\documentclass[
    ,final            
  ]
  {aipproc}

\layoutstyle{8x11single}

\newcommand{\be}{\begin{equation}}
\newcommand{\ee}{\end{equation}}
\newcommand{\ba}{\begin{eqnarray}}
\newcommand{\ea}{\end{eqnarray}}
\def\eps{\epsilon}
\def\veps{\varepsilon}
\def\g{\gamma}
\def\p{\prime}
\def\aprle{\buildrel < \over {_{\sim}}}
\def\aprge{\buildrel > \over {_{\sim}}}

\begin{document}

\title{Observational Consequences of GRBs as Sources of Ultra High
Energy Cosmic Rays}

\classification{98.70.Sa, 98.70.Rz}
\keywords      {gamma rays: bursts---gamma rays: theory---radiation
  mechanisms: nonthermal}

\author{Soebur Razzaque}{
  address={Space Science Division, U.S. Naval Research Laboratory, 
Washington, DC 20375, USA},
altaddress={National Research Council Research Associate}
}

\author{Charles D. Dermer}{
  address={Space Science Division, U.S. Naval Research Laboratory, 
Washington, DC 20375, USA},
}

\author{Justin D. Finke}{
  address={Space Science Division, U.S. Naval Research Laboratory, 
Washington, DC 20375, USA},
altaddress={National Research Council Research Associate}
}

\author{Armen Atoyan}{
  address={Concordia University, Montreal, Quebec H3G 1M8, Canada}
}

\begin{abstract} Gamma-ray bursts (GRBs) have long been considered as
candidates of ultrahigh-energy cosmic rays (UHECRs).  We investigate
the signatures of CR proton acceleration in the GRBs by consistently
taking into account their hadronic and electromagnetic interactions.
We discuss the implications of our findings for high-energy gamma ray
observations with the recently launched {\em Fermi Gamma-ray Space
Telescope.}  \end{abstract}

\maketitle


\section{Introduction}

Gamma-ray bursts are the most powerful explosions in the universe
releasing about $10^{51}$~ergs of energy in keV--MeV $\gamma$-rays
within tens of seconds.  Emission of this non-thermal radiation via
synchrotron and/or Compton scattering requires acceleration of
electrons to very high energies, probably by a Fermi mechanism in
relativistic shocks.  Rapid variability time scales ($t_v
\sim$0.001--1~s) observed in non-thermal $\gamma$-ray light curves
implies that the emission region is compact and moves with a large
 bulk Lorentz factor ($\Gamma \sim$100--1000) towards the observer.

In the context of the fireball shock model (see, e.g.,
Refs.~\cite{Piran05,Meszaros06} for reviews), successively ejected
materials in the form of fireballs from a central engine (black hole
or a magnetar) collide with each other, due to differences in their
bulk Lorentz factors, forming relativistic forward and reverse shocks
(internal shocks).  A baryonic (assumed mostly protons)
contamination/load in an otherwise pure $e^\pm$ and $\gamma$ dominated
fireball or the energy release from the central engine may change in
time and lead to a varying bulk Lorentz factor.  Thus a baryonic
component is required in this model to explain rapid $\gamma$-ray
variability.

Protons can be accelerated to UHE in the same shock which may be
responsible for electron acceleration in the GRB jet~\cite{Waxman95}.
The energy losses and size scale of the shocked region limit, however,
the maximum proton energy.  Synchrotron radiation and photohadronic
($p\gamma$) interactions are dominant energy loss channels for the UHE
protons and the lost energy is converted to high-energy $\gamma$-rays
and
neutrinos~\cite{Waxman97,Rachen98,Dermer03,Razzaque04,Murase06,Asano07,Gupta07,Razzaque09}.
The acceleration of protons to UHE takes place on a longer time scale
than the acceleration of electrons in the shocks.  Therefore, a
characteristic time delay is expected for proton-induced high-energy
$\gamma$-ray and neutrino emission as compared to the keV--MeV
emission by electrons via synchro-Compton mechanism(s).  We discuss
these signatures of UHE cosmic ray acceleration in the GRB jet as
currently being probed by the Large Area Telescope (LAT) and the
Gamma-ray Burst Monitor (GBM) onboard the {\em Fermi Gamma-ray Space
Telescope.}

\section{Proton acceleration and energy losses}

The $\gamma$-ray spectrum in the keV--MeV range is well-fitted by the
Band function~\cite{Band93} which is defined as
\be 
{\cal B} (\veps) = {\cal A} \times \cases{ 
(\veps/100~{\rm keV})^\alpha \exp(-\veps/\veps_0) 
~;~  \veps < (\alpha - \beta)\veps_0 \cr
[(\alpha - \beta)(\veps_0/100~{\rm keV})]^{\alpha-\beta}
\exp(\beta-\alpha) (\veps/100~{\rm keV})^\beta 
~;~ \veps \ge (\alpha - \beta)\veps_0 ~, }
\label{Band_function}
\ee
where $\veps$ is the photon energy measured in keV and ${\cal A}$ is
the normalization factor measured in units of
photons~cm$^{-2}$~s$^{-1}$~keV$^{-1}$ at 100~keV. An integration of
the quantity $\veps {\cal B}(\veps)$ over a given energy band (which
we assumed to be 1~keV -- 1~GeV throughout our calculation) and time
interval gives the fluence ${\cal S}$ (e.g., in units of
ergs~cm$^{-2}$) in that energy and time range. The total
isotropic-equivalent energy release from a burst at redshift $z$ and
luminosity distance $d_L$ is ${\cal E} = 4\pi d_L^2 {\cal S}/(1+z)$.

The proper number density of photons (e.g., in units of
photons~cm$^{-3}$~keV$^{-1}$) in the comoving fireball frame (all
variables are primed in this frame) can be obtained from the Band
function as
\be
n^\p_\gamma (\veps^\p) \simeq \frac{2d_L^2}{\Gamma^4 t_v^2 c^3} 
{\cal B} \left( \frac{\veps^\p \Gamma} {1+z} \right) ~,
\label{comoving_gam_density}
\ee
where $\veps^\p = \veps (1+z)/\Gamma$ and the size of the emitting
region is $R^\p \approx \Gamma t_v c/(1+z)$.  The energy density of
photons in the comoving frame is $U^\p_\gamma = \int_{\veps^\p_{\rm
min}}^{\veps^\p_{\rm max}} \veps^\p n^\p_\gamma (\veps^\p) d\veps^\p$.
The energy density of baryons is assumed to be $U^\p_p =
U^\p_\gamma/\eps_e$, where $\eps_e < 1$, in the fast-cooling scenario.
The magnetic energy density is $U^\p_B = B^{\p 2}/8\pi = \eps_B
U^\p_p$, and the magnetic field is $B^\p = \sqrt{8\pi
(\eps_B/\eps_e)U^\p_\gamma}$ with $\eps_B < 1$.  We choose the typical
Band function parameters $\alpha=-1$, $\beta=-2.2$, $\eps_0 = 1$~MeV,
${\cal A} = 10^{-2}$~ph~cm$^{-2}$~keV$^{-1}$ and $z=2$ (corresponding
to $d_L \simeq 4.85\times 10^{28}$~cm for the standard $\Lambda$CDM
cosmology), and we have listed in Table~\ref{tab:a} different derived
quantities for $t_v = 10^{-3}$--1~s and for 1~keV $\le \veps \le$
1~GeV.  We assumed that $\Gamma$ is the same as $\Gamma_{\rm min}$
which has been derived using the condition that the fireball becomes
optically thin to $\g\g \to e^\pm$ pair production at 1~GeV (see,
e.g., Refs.~\cite{Lithwick01,Razzaque04,Dermer05}).

\begin{table}
\begin{tabular}{lccrrr}
\hline
\tablehead{1}{r}{b}{$t_v$~(s)}
  & \tablehead{1}{c}{b}{${\cal S}$~(ergs~cm$^{-2}$)}
  & \tablehead{1}{c}{b}{${\cal E}$~(ergs)}
  & \tablehead{1}{r}{b}{$\Gamma$}
  & \tablehead{1}{r}{b}{$R = \Gamma R^\p$~(cm)}
  & \tablehead{1}{r}{b}{$B^\p\sqrt{\eps_B/\eps_e}$ (G)} \\
\hline
$10^0$ & $3.2\times 10^{-5}$ & $3.2\times 10^{53}$ & 385 
  & $1.5\times 10^{15}$ & $6.3\times 10^3$ \\
$10^{-1}$ & -- & -- & 550 & $3.0\times 10^{14}$ & $2.2\times 10^4$ \\
$10^{-2}$ & -- & -- & 790 & $6.2\times 10^{13}$ & $7.3\times 10^4$ \\
$10^{-3}$ & -- & -- & 1135 & $1.3\times 10^{13}$ & $2.5\times 10^5$ \\
\hline
\end{tabular}
\caption{Derived quantities from Band function parameters: 
$\alpha = -1$, $\beta = -2.2$, $\eps_0 = 10^3$~keV, 
${\cal A} = 10^{-2}$~ph~cm$^{-2}$~keV$^{-1}$ at 100~keV, and 
$z=2$ for a 10~s long GRB}
\label{tab:a}
\end{table}

The acceleration time scale for protons in the electric field induced
by $B^\p$ is proportional to the Larmor time ($E^\p_p/eB^\p$) and is
given by
\be 
t^\p_{p,\rm acc} = \frac{\phi \hbar}{m_p^2 c^4} \frac{B_{\rm cr}}{B^\p}
E^\p_{p} \simeq 11~\frac{\phi E^\p_{p,9}}{B^{\p}_{4}}~{\rm s} ~, 
\label{p_acc_time} 
\ee
where $B_{\rm cr} = m_p^2c^3/e\hbar = 1.488\times 10^{20}$~G, $B^\p =
10^4 B^\p_{4}$~G, $E^\p_{p} = 10^9 E^\p_{p,9}$~GeV and $\phi \aprge 1$ 
is the number of gyro-radii required to increase a particle
energy by a factor of 2.7.  The synchrotron cooling time for protons
in the same magnetic field is
\be 
t^\p_{p,\rm syn} = \frac{9}{4} \frac{\hbar^2}{r_e m_e c}
\frac{B_{\rm cr}^2}{B^{\p 2}} 
\frac{1}{E^\p_{p}} 
\simeq \frac{45}{B^{\p 2}_{4} E^\p_{p,9}}~{\rm s} ~.
\label{p_sync_time} 
\ee
The photohadronic cooling time scale, which does not depend on the
magnetic field but on the observed keV--MeV $\gamma$-ray source
density in equation~(\ref{comoving_gam_density}), is given by
\be
t^{\p -1}_{p\gamma} = 
\frac{m_p^2 c^5}{2E_p^{\p 2}} 
\int_0^\infty d\veps^\p \frac{n^\p_\gamma(\veps^\p)}{\veps^{\p 2}}
\int_0^{2E^\p_p\veps^\p/m_pc^2} d\veps^\p_r
\sigma_{p\gamma} (\veps^\p_r) f(\veps^\p_r) \veps^\p_r 
\simeq c\sigma_0 \int_{\veps^\p_{\rm th} m_pc^2/2E^\p_p}^\infty
d\veps^\p n^\p_\gamma(\veps^\p)
\label{pg_cool_time}
\ee
for single $p\g$ scattering.  We approximated a $p\g$ cross-section of
$\sigma_0 \equiv \sigma_{p\gamma} (\veps^\p_r) f(\veps^\p_r) =
68$~$\mu$b for $\veps^\p_r \ge \veps^\p_{\rm th}$ following
Ref.~\cite{Dermer07}.  Here $\veps^\p_r = \veps^\p (1-\beta_p
\cos\theta)E^\p_p/m_pc^2$ is the photon energy evaluated in the
proton's rest frame for the angle $\theta$ between the directions of
the proton and target photon, and we take $\veps^\p_{\rm th} \simeq
0.2$~GeV as the threshold photon energy for pion production in the
rest frame of the proton.

\begin{figure} \includegraphics[height=.45\textheight]{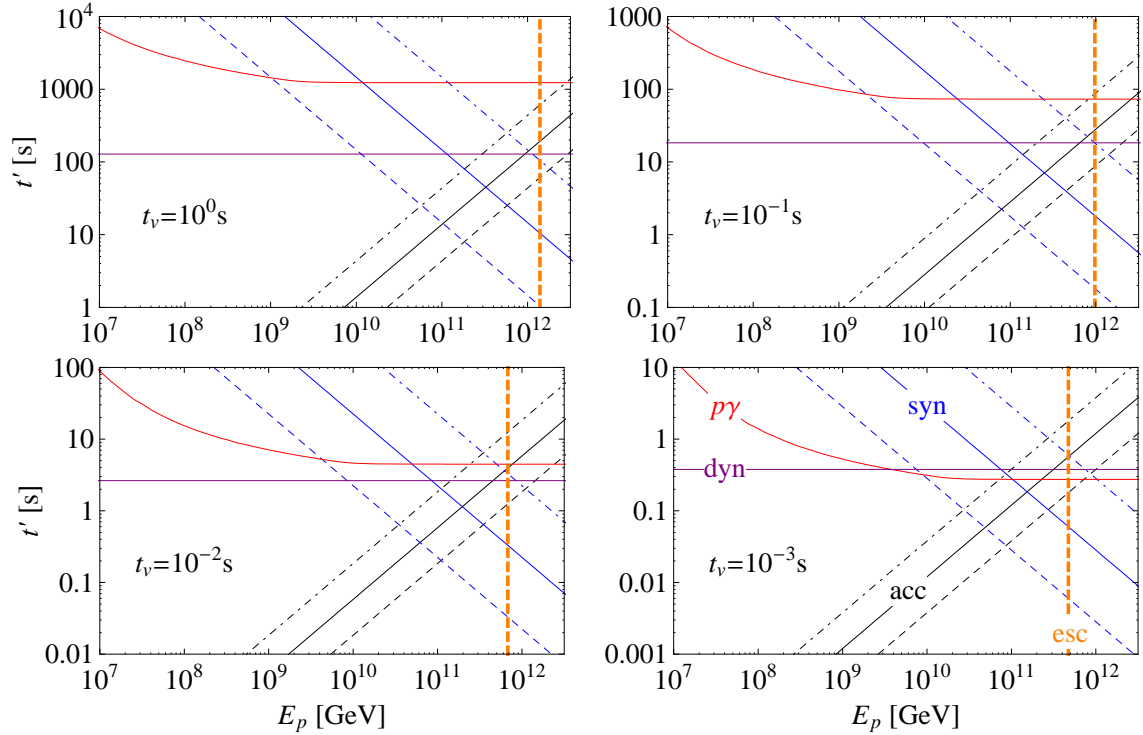}
\caption{The acceleration and cooling times (labeled in the
lower-right panel) for protons for the GRB parameters assumed in
Table~\ref{tab:a} with different $t_v$.  The dashed, solid and
dot-dashed curves in the sets labeled ``acc'' and ``syn'' are for
$\eps_B/\eps_e$ = 0.1, 1 and 10 respectively.  We used the Band
function in equation~(\ref{Band_function}) in the 1~keV -- 1~GeV
observed photon energy range with derived parameters given in
Table~\ref{tab:a}, and $\phi =1$ in equation~(\ref{p_acc_time}) for
our calculation.  The maximum proton energy can be read off from a
plot when the ``acc'' curve, for a given $B^\p$ field, is intersected
by one of the cooling curves ``$p\gamma$'' or ``syn'' for the same
$B^\p$ field, or by the ``dyn'' curve.  The top left panel for $t_v =
1$~s and $\eps_B/\eps_e$ = 1 correspond to $E_{p,\rm max} \simeq
3\times 10^{20}$~eV, the highest among all choices of the parameters
$t_v$ and $\eps_B/\eps_e$. See main text for details.}
\label{fig:times} \end{figure}

We have plotted in Fig.~\ref{fig:times} different time scales for
protons with the assumed Band function parameters, redshift and four
different variability time scales (see Table~\ref{tab:a}).  Two sets
of three straight lines, with positive and negative slopes, correspond
to acceleration and synchrotron cooling times respectively, for
$\eps_B/\eps_e =$10 (dashed lines), 1 (solid lines) and 0.1
(dot-dashed lines) in Table~\ref{tab:a}.  The horizontal solid line in
each panel correspond to the dynamic time $t^\p_{\rm dyn} \approx t_v
\Gamma/(1+z)$. The vertical thick-dashed line is the proton escape
time $t^\p_{\rm esc} = (3/2) (m_p^2 c^4/\hbar) (B^\p/B_{\rm cr}) t^{\p
2}_{\rm dyn}/E^\p_p$ in the Bohm diffusion limit, plotted for the
proton energy at which $t^\p_{\rm esc} = t^\p_{\rm dyn}$. Note that
for all four $t_v$ plotted in Fig.~\ref{fig:times}, synchrotron
cooling effectively determines the maximum proton energy if
$\eps_B/\eps_e \aprge 1$.  The photohadronic cooling (smoothly-curved
solid line labeled $p\gamma$) is important for highly variable GRBs as
the internal photon density [see
equation~(\ref{comoving_gam_density})] increases to allow frequent
interactions by high-energy protons.  The protons may not escape the
shocked region for the parameters assumed here.

Proton-synchrotron radiation and photohadronic interactions (via
$\pi^0$ decays) produce very high-energy photons.  Photohadronic
interactions also produce very high-energy electrons (via $\pi^+$
decays).  These particles initiate electromagnetic cascades in the GRB
fireball.  We postpone the details of the cascade modeling and
emission for future work.  In the next section we concentrate on an
analyic estimate of the proton-synchrotron radiation, which dominates
for $\eps_B/\eps_e \aprge 1$ in the GRB fireball.  Note that, in this
limit the Compton parameter $Y= (-1 + \sqrt{1+4\eps_e/\eps_B})/2
\aprle 1$ in the fast-cooling scenario and the Compton scattering of
synchrotron photons to high-energy $\g$-rays (SSC) may not be
important~\cite{Sari01}.

\section{Proton synchrotron radiation}

\begin{figure} \includegraphics[height=.25\textheight]{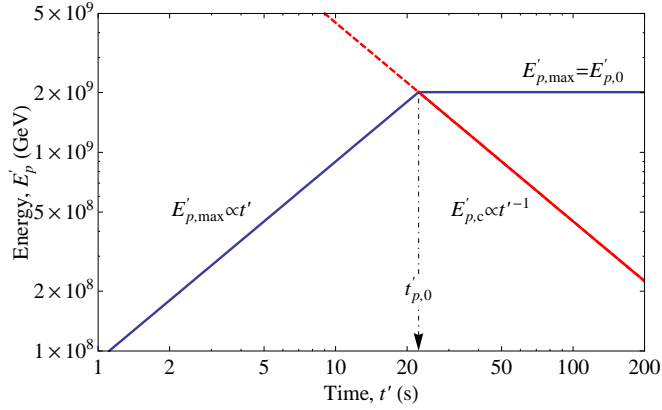}
\caption{Maximum shock-accelerated energy ($E^\p_{p,max}$) and
characteristic synchrotron cooling energy ($E^\p_{p,c}$) for protons as
functions of the comoving time.  Note that the synchrotron cooling is
important for $t^\p \ge t^\p_{p,0}$ where $t^\p_{p,0}$ is denoted by
the arrow.}  \label{fig:E_crits} \end{figure}

Assuming that synchrotron radiation limits proton acceleration, the
maximum shock-accelerated proton energy as a function of time in the
shocked fluid frame is
\be 
E^\p_{p,\rm max}(t^\p) = 
\cases{ \frac{m_p^2 c^4}{\phi\hbar}
\frac{B^\p}{B_{\rm cr}} t^\p ~;~ 0 \le t^\p < t^\p_{p,0} \cr
\left( \frac{9}{4} \frac{\hbar m_p^2 c^3} {\phi r_e m_e}
\frac{B_{\rm cr}}{B^\p}\right)^{1/2} ~;~ t^\p_{\rm dyn} \ge t^\p \ge
t^\p_{p,0} } 
\label{p_max_E} 
\ee
The maximum proton energy in the second case in
equation~(\ref{p_max_E}) is found from the condition $t^\p_{p,\rm acc}
= t^\p_{p,\rm syn}$ and the time to accelerate protons to this maximum
energy are given by
\ba
E^\p_{p,0} &=& \left( \frac{9}{4} \frac{\hbar m_p^2 c^3} {\phi r_e m_e}
\frac{B_{\rm cr}}{B^\p}\right)^{1/2} 
\approx \frac{2\times 10^9}{\phi^{1/2} B^{\p 1/2}_{4}}~{\rm GeV, ~and} 
\label{maximum_E} \\
t^\p_{p,0} &=& \left( \frac{9}{4} \frac{\phi \hbar^3
B_{\rm cr}^3}{r_e m_e m_p^2 c^5 B^{\p 3}} \right)^{1/2} \approx
22 ~\frac{\phi^{1/2}}{B^{\p 3/2}_{4}} ~{\rm s},
\label{max_E_time}
\ea
respectively.  Thus for $t^\p \ge t^\p_{p,0}$ the maximum proton
energy remains constant at $E^\p_{p,0}$ (see Fig.~\ref{fig:E_crits}).
The synchrotron cooling, however, becomes significant with increasing
time.  The characteristic synchrotron cooling-break energy for protons
from the condition $t^\p_{p,\rm syn} = t^\p$, and for $t^\p_{\rm dyn}
\ge t^\p \ge t^\p_{p,0}$ is given by (see Fig.~\ref{fig:E_crits})
\be
E^\p_{p,\rm c} (t^\p) = 
\frac{9}{4} \frac{\hbar^2 B_{\rm cr}^2} {r_e m_e
c B^{\p 2}} \frac{1}{t^\p} \approx  
\frac{5\times 10^8}{B^{\p 2}_{4} t^\p_2} ~{\rm GeV},
\label{p_sync_char_E} 
\ee
where $t^\p_2 = t^\p/100$~s.  Protons above this energy cool
efficiently down to $E^\p_{p,\rm c} (t^\p)$ within $t^\p$.  As a
result the proton spectrum can be approximated by a broken power law
with the spectral index softened by unity for $E^\p_{p} > E^\p_{p,\rm
c}$~\cite{Sari98}.

\begin{figure} \includegraphics[height=.25\textheight]{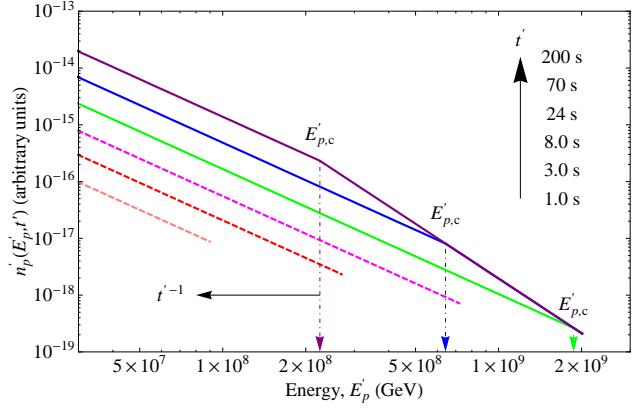}
\caption{ Shock-accelerated proton spectrum at increasing comoving
time (from bottom to top).  Total number of protons swept-up by the
shock increases linearly with time as in
equation~(\ref{p_spectrum_t}).  At early times, $t^\p < t^\p_{p,0}$ in
equations~(\ref{p_max_E}) and (\ref{max_E_time}), the maximum proton
energy increases linearly (dashed lines) and saturates to a constant
value $E^\p_{p,0}$ given in equation~(\ref{maximum_E}) for $t^\p \ge
t^\p_{p,0}$ (solid lines).  As the synchrotron cooling becomes
important for $t^\p \ge t^\p_{p,0}$, the spectrum shows a break at
$E^\p_{p,\rm c}$ given in equation~(\ref{p_sync_char_E}) which
decreases with time as $\propto t^{\p -1}$.  The spectrum is $\propto
E^{\p -\kappa}$ below $E^\p_{p,\rm c}$ and $\propto E^{\p -\kappa-1}$
above $E^\p_{p,\rm c}$. } \label{fig:E_spect} \end{figure}

The number density of particles behind the shock is $4n^\p \Gamma_{\rm
rel}$, where $n^\p$ is the pre-shock number density in the GRB
fireball.  As the shock crosses the fireball, it sweeps-up particles
and the number density of shocked particles increases $\propto t^\p$.
We assume that a time-dependent fraction of the shock energy density
$4n^\p m_pc^2\Gamma_{\rm rel}^2$ is channeled to the high energy
protons. In the most optimistic scenario, the time-dependent proton
number spectrum can be written as
\be
n^\p_p (E^\p_p, t^\p) \approx  4n^\p \frac{t^\p}{t^\p_{\rm dyn}} C_p
\times \cases{ (E^\p_p/E^\p_{p,\rm min})^{-\kappa} ~;~
E^\p_{p,\rm min} \le E^\p_p < E^\p_{p,\rm c}(t^\p) \cr
(E^\p_{p,c}/E^\p_{p,\rm min}) (E^\p_p/E^\p_{p,\rm min})^{-\kappa-1} 
~;~ E^\p_{p,\rm c}(t^\p) \le E^\p_p \le E^\p_{p,\rm max}(t^\p)} 
\label{p_spectrum_t}
\ee
where $C_p$ is an overall normalization factor and $E^\p_{p,\rm min}
\sim \Gamma_{\rm rel} m_pc^2$ is the minimum proton energy, and
$\Gamma_{\rm rel} \sim 1$--10 is the relative Lorentz factors between
two colliding shells. The snapshot of this spectrum is plotted in
Fig.~\ref{fig:E_spect} at different $t^\p$, and for arbitrarily fixed
$n^\p$ and $C_p$.

\begin{figure}
\includegraphics[height=.25\textheight]{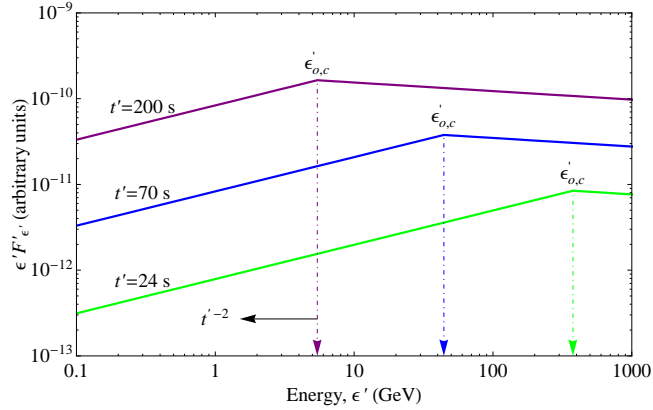} \caption{
Proton synchrotron radiation spectrum at three different times $t^\p >
\veps^\p_{o,\rm c}$.  Note that the peak of the synchrotron radiation
spectrum $\veps^\p_{o,\rm c}$ decreases with time as $t^{\p -2}$ as
indicated by the arrows.  The spectrum is $\propto \veps^{\p
-(\kappa-1)/2}$ below $\veps^\p_{o,\rm c}$ and $\propto \veps^{\p
-\kappa/2}$ above $\veps^\p_{o,\rm c}$, where $\kappa \aprge 2$ is the
spectral index for shock-accelerated protons.}
\label{fig:sync_spectrum} \end{figure}

The synchrotron power and the typical synchrotron photon energy
emitted by a proton of energy $E^\p_{p}$ are given by
\ba 
P^\p(E^\p_p) &=& \frac{4}{9} \frac{r_e m_e c B^{\p 2}} {\hbar^2
B_{\rm cr}^2} E^{\p 2}_p, ~{\rm and} \label{syn_power_per_p} \\
\veps^\p_{o} &=& \frac{3}{2} \frac{B^\p}{B_{\rm cr}} 
\frac{E^{\p 2}_p} {m_pc^2} 
\label{typical_syn_photon_E} ~,
\ea
respectively.  For the characteristic proton synchrotron cooling
energy in equation~(\ref{p_sync_char_E}), the typical synchrotron
photon energy from equation~(\ref{typical_syn_photon_E}) for
$t^\p_{\rm dyn} \ge t^\p \ge t^\p_{p, 0}$ is
\be
\veps^\p_{o,\rm c} (t^\p) = \frac{243}{32} \frac{\hbar^4 B_{\rm cr}^3}
{r_e^2 m_e^2 m_p c^4 B^{\p 3}} \frac{1}{t^{\p2}} 
\approx  \frac{22}{B^{\p 3}_{4} t_2^{\p 2}} ~{\rm GeV}.
\label{typ_syn_phot_E_char_Ep} 
\ee
The ratio of the synchrotron power to the typical photon energy
roughly gives the spectral power. The total spectral emissivity (e.g.,
in units of ergs~cm$^{-3}$~s$^{-1}$~GeV$^{-1}$) from all protons with
energy $E^\p_{p,\rm c}$ from the distribution in
equation~(\ref{p_spectrum_t}) is given by
\be 
{\cal P}^\p_{\veps^\p_{o,\rm c}} (t^\p) \approx 
\frac{P^\p(E^\p_{p,\rm c}(t^\p))} {\veps^\p_{o,\rm c}(t^\p)}
\times n^\p_p (E^\p_{p,\rm c},t^\p) 
\approx  \frac{8}{27} \frac{r_e m_e m_p c^3 B^{\p}}
{\hbar^2 B_{\rm cr}} n^\p_p (E^\p_{p,\rm c},t^\p) .
\label{syn_power_cool_break}
\ee
The corresponding time-dependent energy flux (e.g.,
in units of ergs~cm$^{-2}$~s$^{-1}$) of the proton synchrotron
radiation, in analogy with electron synchrotron radiation in the
slow-cooling regime~\cite{Sari98}, is
\be 
{\cal F}^\p_{\veps^\p} (t^\p) \propto E^\p_{p,\rm c} {\cal
P}^\p_{\veps^\p_{o,\rm c}} (t^\p) \times \cases{
(\veps^\p/\veps^\p_{\rm min})^{1/3} 
~;~ \veps^\p < \veps^\p_{\rm min} \cr
(\veps^\p/\veps^\p_{\rm min})^{-(\kappa-1)/2} 
~;~ \veps^\p_{\rm min} \le \veps^\p \le \veps^\p_{o,\rm c} (t^\p) \cr 
(\veps^\p_{o,\rm c}/\veps^\p_{\rm min})^{-(\kappa-1)/2}
(\veps^\p/\veps^\p_{o,\rm c})^{-\kappa/2} 
~;~ \veps^\p_{o,\rm c}(t^\p) \le \veps^\p \le 
\veps^\p_{\rm max} (t^\p) . } 
\label{p_syn_spectrum} 
\ee
Here $\veps^\p_{\rm min}$ and $\veps^\p_{\rm max}$ are respectively
the typical photon energies for the protons with energies $E^\p_{p,\rm
min}$ and $E^\p_{p,\rm max}$.  We have plotted this proton synchrotron
radiation spectrum at different times $t^\p \ge t^\p_{p,0}$ in
Fig.~\ref{fig:sync_spectrum}.  High energy photons $\veps^\p
\aprge$ MeV, however, produce $e^\pm$ pairs which initiate cascades.
Note that the proton synchrotron radiation and the cascade radiation
build-up gradually with time as the shock propagates through the GRB
fireball.  As a result, high-energy emission from proton synchrotron
radiation would be delayed compared to the prompt keV--MeV emission
from shock-accelerated primary electrons.

\section{Discussion}

We have investigated different scenarios of fireball shock model for
GRBs in which protons can be accelerated to ultrahigh energies and
produce observable $\gamma$-ray signatures.  We found that --- for the
Band function parameters adopted here, in the 1 keV -- 1 GeV range,
typical of a bright long duration GRB --- synchrotron radiation is the
dominating energy loss channel for UHE protons for the model
parameters $\eps_B/\eps_e \aprge 1$, the acceleration parameter $\phi
\approx 1$ and for the variability time scale $t_v \aprge 10^{-3}$~s.
An SSC component of high-energy $\g$-rays, which are expected to form
simultaneously with keV--MeV synchrotron photons, is suppressed in the
case $\eps_B/\eps_e \aprge 1$ as well.  Photohadronic ($p\g$)
interactions can dominate the energy losses by UHE protons if the GRB
fireball is highly compact with $\sim$1~ms scale variability or more
target soft photons in the fireball than considered here, and for
$\eps_B/\eps_e < 1$.  In this case an SSC component is expected.  We
also found that proton synchrotron radiation and associated
electromagnetic cascade emission, from synchrotron photons or/and
pion-decay secondaries from $p\g$ interactions, can produce
high-energy $\g$-rays which are delayed compared to the prompt
keV--MeV emission.  The delay is due to the time required for protons
to be accelerated to UHE when they suffer large energy losses.


\begin{theacknowledgments}
Work supported by the Office of Naval Research and NASA grants.
\end{theacknowledgments}


\begin{thebibliography}{99}

\bibitem{Piran05} T.~Piran, \emph{Rev. Mod. Phys.}, \textbf{76}, 1143
(2005).

\bibitem{Meszaros06} P.~M{\'e}sz{\'a}ros, \emph{Rep. Prog. Phys.},
\textbf{69}, 2259 (2006).

\bibitem{Waxman95} E.~Waxman, \emph{Phys. Rev. Lett.}, \textbf{75},
386 (1995).

\bibitem{Waxman97} E.~Waxman, and J.~N. Bahcall,
\emph{Phys. Rev. Lett.}, \textbf{78}, 2292 (1997).

\bibitem{Rachen98} J.~P. Rachen, and P.~M{\'e}sz{\'a}ros,
\emph{Phys. Rev. D}, \textbf{58}, 123005 (1998).

\bibitem{Dermer03} C.~D. Dermer, and A.~Atoyan,
\emph{Phys. Rev. Lett.}, \textbf{91}, 071102 (2003).

\bibitem{Razzaque04} S.~Razzaque, P.~M{\'e}sz{\'a}ros, and B.~Zhang,
\emph{Astrophys. J.}, \textbf{613}, 1072 (2004).

\bibitem{Murase06} K.~Murase, and S.~Nagataki, \emph{Phys. Rev. D},
\textbf{73}, 063002 (2006).

\bibitem{Asano07} K.~Asano, and S.~Inoue, \emph{Astrophys. J.},
\textbf{671}, 645 (2007).

\bibitem{Gupta07} N.~Gupta, and B.~Zhang, \emph{MNRAS}, \textbf{380},
78 (2007).

\bibitem{Razzaque09} S.~Razzaque, O.~Mena, and C.~D. Dermer,
\emph{Astrophys. J. Lett.}, \textbf{691}, L37 (2009).

\bibitem{Band93} D.~Band et al. \emph{Astrophys. J.}, \textbf{413},
281 (1993).

\bibitem{Lithwick01} Y.~Lithwick, and R.~Sari, \emph{Astrophys. J.},
\textbf{555}, 540 (2001).

\bibitem{Dermer05} C.~D. Dermer, ``{Photoabsorption of Gamma Rays in
Astrophysical Jets}'' in \emph{The Tenth Marcel Grossmann Meeting},
edited by M.~Novello et al., World Scientific, Singapore, 2005,
pp. 1385.

\bibitem{Dermer07} C.~D. Dermer, \emph{Astrophys. J.}, \textbf{664},
384 (2007).

\bibitem{Sari01} R.~Sari, and A.~Esin, \emph{Astrophys. J.},
\textbf{548}, 787 (2001).

\bibitem{Sari98} R.~Sari, T.~Piran, and R.~Narayan,
\emph{Astrophys. J. Lett.}, \textbf{497}, L17 (1998).

\end{thebibliography}
\end{document}